\documentclass[runningheads]{llncs}
\usepackage{graphicx}
\begin{document}
\title{Patient-centric health data sovereignty: an approach using Proxy re-encryption\thanks{This work was partially supported by the Norte Portugal Regional Operational Programme (NORTE 2020), under the PORTUGAL 2020 Partnership Agreement, through the European Regional Development Fund (ERDF), within project ``Cybers SeC IP'' (NORTE-01-0145-FEDER-000044).}}
\titlerunning{Patient-centric health data sovereignty: an approach using PRE}
\author{Bruno Rodrigues\inst{1}\orcidID{0009-0001-2929-0836} \and
 Ivone Amorim\inst{2}\orcidID{0000-0001-6102-6165} \and
Ivan Silva\inst{2}\orcidID{0009-0009-8480-9352} \and Alexandra Mendes\inst{3}\orcidID{0000-0001-8060-5920}}
\authorrunning{B. Rodrigues et al.}
\institute{Faculty of Engineering, University of Porto, Porto, Portugal \\ \email{up202103390@edu.fe.up.pt}
\and PORTIC – Porto Research, Technology \& Innovation Center, Polytechnic of Porto~(IPP), 4200-374 Porto, Portugal \\
\email{ivone.amorim@portic.ipp.pt}
\and Faculty of Engineering, University of Porto, Porto, Portugal \& HASLab/INESC TEC \\ \email{afmendes@fe.up.pt}
}
\maketitle              % typeset the header of the contribution
\begin{abstract}

The exponential growth in the digitisation of services implies the handling and storage of large volumes of data. Businesses and services see data sharing and crossing as an opportunity to improve and produce new business opportunities. The health sector is one area where this proves to be true, enabling better and more innovative treatments. Notwithstanding, this raises concerns regarding personal data being treated and processed. In this paper, we present a patient-centric platform for the secure sharing of health records by shifting the control over the data to the patient, therefore, providing a step further towards data sovereignty. Data sharing is performed only with the consent of the patient, allowing it to revoke access at any given time. Furthermore, we also provide a \emph{break-glass} approach, resorting to Proxy Re-encryption (PRE) and the concept of a centralised trusted entity that possesses instant access to patients' medical records. Lastly, an analysis is made to assess the performance of the platform's key operations, and the impact that a PRE scheme has on those operations.

\end{abstract}

\section{Introduction}

The ever growing digitisation of services that we use daily, as well as the increasing interest in data crossing and sharing to improve processes, services, and achieve new business opportunities, raises concerns regarding how data is handled and processed. In the healthcare sector, data sharing is not only beneficial, but also needed to provide the best care possible to the patients. However, this data is also highly sensitive, which requires special care. Several governmental measures have already been taken to improve and standardise the way in which data is shared, such as the European Data Governance Act~\cite{edga_2022}, GDPR\footnote{https://data.europa.eu/eli/reg/2016/679/oj}, and, more specifically in personal health information, HIPAA\footnote{https://www.cdc.gov/phlp/publications/topic/hipaa.html} and HITECH\footnote{https://www.hipaajournal.com/what-is-the-hitech-act/}. These directives instigate a user-centric paradigm, granting individuals sovereignty over their data.

Several approaches have been proposed for ensuring security and privacy in e-health systems. Conventional encryption techniques like AES and ECC are commonly used~\cite{8747355}. However, these techniques become problematic when data needs to be shared among multiple entities due to redundancy and computational burden~\cite{enisa}. Attribute-Based Encryption (ABE) is another solution~\cite{enisa}, but it has its own complexities and limitations, such as managing attribute-based keys and overriding policies in emergencies~\cite{FERNANDEZALEMAN2013541}. ABE also lacks the fine-grained access control necessary for a patient-centric sovereign approach.

Proxy Re-encryption (PRE) is a cryptographic solution for secure data sharing without prior knowledge of the recipient. Unlike ABE, it does not rely on policies or attributes. PRE converts a ciphertext to a recipient's key without revealing the plaintext to the intermediary entity. It is particularly useful in semi-trusted cloud environments~\cite{NUNEZ2017193}. In e-health, PRE has already been used to securely share medical records~\cite{app12094353,THILAKANATHAN2014102,8319175,YUKSEL20171}, including in emergency scenarios~\cite{8319175}. However, challenges remain in terms of revocability, computational effort, and safeguarding emergencies~\cite{YUKSEL20171}. Existing solutions for emergency scenarios are limited and rely on assumptions that may impact efficiency and reliability.

In this context, it is necessary to develop a platform that addresses the aforementioned concerns. This includes enabling more control over the data by the patient while ensuring the safety of that data, even in semi-trusted environments. This contributes to the collaborative aspect of e-health and thus enables better treatments and advancements in the health sector.

In this paper, we present a platform that leverages PRE to enhance health data sharing. Umbral's PRE~\cite{Umbral} is used as the foundation for re-encryption processes, through which we achieve unidirectionality and non-interactivity, ensuring secure re-encryption from the patient to the data user (e.g., practitioners or health centres) without requiring the data user's private key. This approach centres on the expressed opinion of the patient to authorise data sharing, eliminating the need for prior identification of authorised parties — a drawback identified in previous solutions. Additionally, our platform offers revocability options, such as time-based access limits and patient-initiated access revocation. Importantly, the revocation of accesses does not require changes to the encrypted healthcare database, distinguishing our platform from the ones that rely on identity and attribute-based PRE schemes.

Furthermore, in the context of healthcare, it is crucial to ensure data sharing in emergency situations when explicit patient consent may not be possible. Our platform addresses this challenge by incorporating a trusted entity for data access when patient authorisation is infeasible.

In summary, our main contributions are: 

\begin{itemize} 
\item A patient-centric platform, that empowers patients with sovereign control over their health data, enabling granular access control and facilitating the sharing of health records only with explicit consent.

\item Robust data protection using Umbral's PRE, ensuring secure and encrypted health data sharing without compromising the data user's private key.

\item A robust access revocation mechanism that enables time-based access limits and supports manual revocation by the patient at any time and with immediate effect.

\item A break-glass mechanism to ensure seamless emergency data access.

\end{itemize}

The remainder of this paper is organised as follows. Section~\ref{sc: pre} introduces basic concepts and definitions, as well as the classification and properties of PRE schemes. Furthermore, an analysis is made concerning the framework on which the access delegation mechanism is based. Section~\ref{sc: related-work} presents the current picture of the PRE and the advancements regarding break-glass scenarios. Section~\ref{sc: data-sovereignty} details the proposed solution and its implementation. Section~\ref{sc: performance-analysis} is concerned with the performance test, respective results and discussion. Lastly, Section~\ref{sc: conclusion} presents the conclusions and future work.

\section{Proxy Re-encryption} \label{sc: pre}

PRE is a cryptographic technique that enables a third-party entity, named proxy, to delegate access to encrypted data, without being able to infer the plaintext content of that data. This is achieved by transforming a ciphertext encrypted under one key into a ciphertext encrypted under a different key.

\subsection{Syntax and basic definitions} \label{ssc: syntax-definitions}

Since PRE can be seen as a way to delegate decryption rights to a party, it is possible to categorise the different entities according to the delegation relation they possess with each other. The \emph{delegator} is the entity that owns the data and delegates decryption rights. The \emph{proxy} is the intermediary entity in the delegation process, which uses a re-encryption key (PRK) to transform the ciphertext encrypted under the delegator's public key into a ciphertext that can be decrypted only by using the delegatee's private key. Finally, the \emph{delegatee} is the entity that accesses the information through delegation of decryption rights by the delegator.

\begin{definition}[PRE]
A PRE scheme can be defined based on five different algorithms:

\begin{itemize} 
    \item \textbf{KeyGen} ---  On input of a security parameter $n$, the key generation algorithm KeyGen outputs a public/private key pair ($pk_A$, $sk_A$) for a given user A.
    
    \item \textbf{ReKey} --- On input of a public/private key pair ($pk_A$, $sk_A$) for user A and a public/private key pair ($pk_B$, $sk_B$) for user B, a PRK $rk_{\; A \rightarrow B}$ is computed.
    
    \item \textbf{Encrypt} --- Given the input of a public key $pk_A$ and a message $m \in {M}$, the encryption algorithm outputs a ciphertext $c_A \in C_1$.
    
    \item \textbf{ReEncrypt} --- On input of a ciphertext $c_A \in C_1$ and a PRK $rk_{\; A \rightarrow B}$, the re-encryption algorithm ReEncrypt transforms a ciphertext $c_A \in C_1$ into a ciphertext $c_B \in C_2$.
    
    \item \textbf{Decrypt} --- Given a private key $sk_A$ from user A and a ciphertext $c_A \, \in \, C_S \;(S \in \{1,2\})$ from user A, the same executes the decryption algorithm and outputs the original message $m \in {M}$.
\end{itemize}
    
\end{definition}

According to Qin et al.\cite{7448446}, a PRE scheme can be classified based on its abilities. For example, regarding its directionality,  we say that the scheme is \emph{unidirectional} if it enables the delegator's ciphertext to be re-encrypted into the delegatee's ciphertext but not vice versa. Otherwise, we call it \emph{bidirectional}. The multi-use/single-use classification focuses on the number of times the PRK can be used to re-encrypt data. In \emph{multi-use} schemes, the PRK can be utilised to perform several re-encryptions. In the case of a \emph{single-use} scheme, the PRK can only be used to perform a single transformation. Interactivity dictates whether the re-encryption is computed using just the public key from the delegatee (\emph{non-interactive} scheme) or both the public and private keys (\emph{interactive} scheme). Depending on the scenario of utilisation, some properties may be more desirable than others.

Other authors classify PRE schemes according to their way of functioning~\cite{Inbarani2013,KF16}. For example, an Identity-Based PRE (IB-PRE) scheme derives public keys from identity attributes (e.g. email). The messages are encrypted using an identity string from the delegatee. Attribute-Based PRE (AB-PRE) schemes allow transforming a ciphertext defined by a set of attributes or access policies into another ciphertext with a different set of attributes. 

\subsection{Umbral's PRE scheme}
The Umbral PRE scheme is, in its essence, a threshold PRE scheme. This scheme features an Elliptic Curve Integrated Encryption Scheme (EICS-KEM) inspired in~\cite{ansi-x9.63} and proposes several improvements over the original PRE scheme proposed by~\cite{blaze1998divertible},  namely unidirectionality, non-interactivity, and verifiability. It relies on the concept of semi-trusted proxies, also known as ursulas. Being a threshold PRE scheme, it splits the PRK according to shares. The \emph{threshold} portion of the scheme dictates the minimum number of those shares required to decrypt the information.

Splitting the PRK across multiple proxies brings some benefits namely eliminating a single point of failure, in case of a malfunction or compromise of one of the proxies the PRK is still safeguarded.

The re-encryption processes in our platform are supported by pyUmbral~\cite{pyumbral}, a Python-based implementation of Umbral.

\begin{figure}[h!]
    \includegraphics[width=\textwidth]{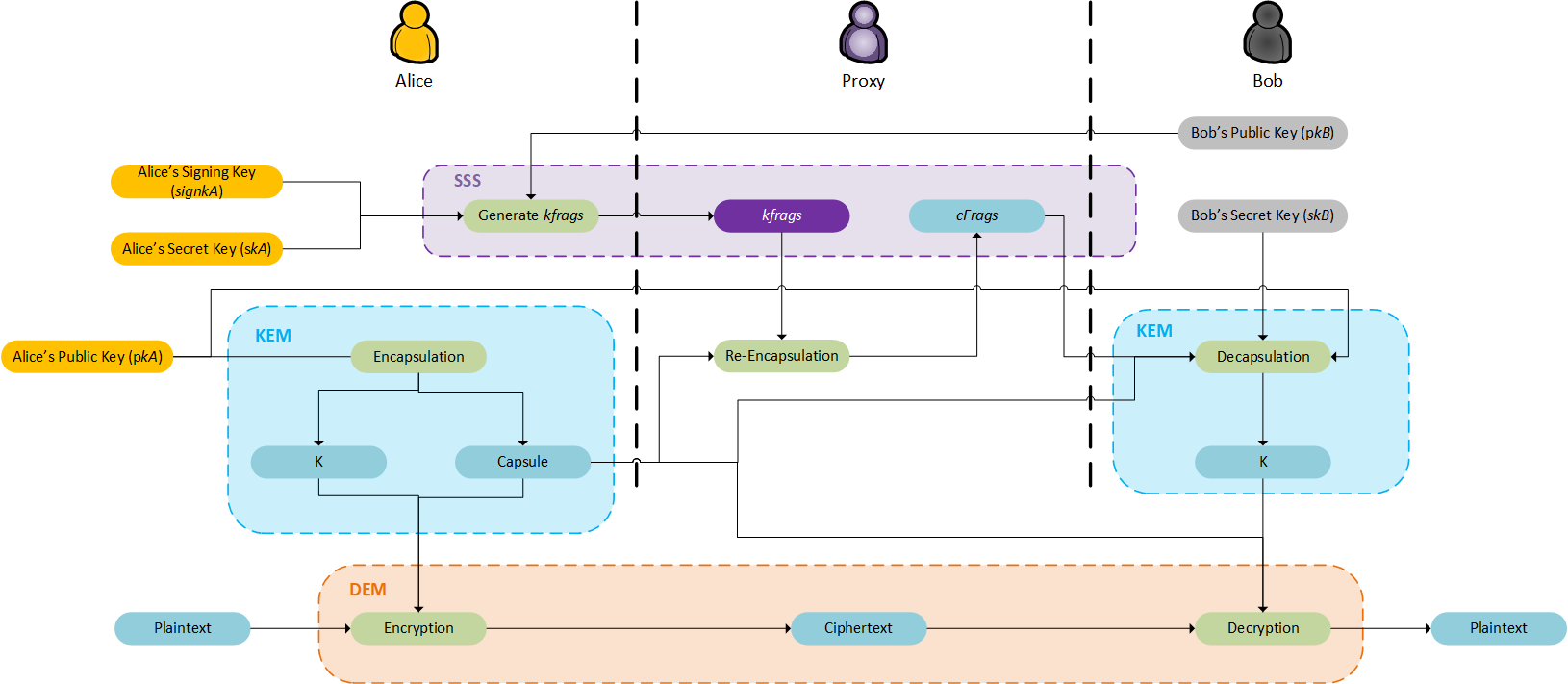}
    \caption{Procedural overview of pyUmbral PRE scheme}
    \label{fig: umbral-overall-flow}
\end{figure}

Fig.~\ref{fig: umbral-overall-flow} presents an overview of the key processes and data flows involved in the Umbral PRE scheme. This system beholds seven main processes: \emph{Encapsulation}, \emph{Encryption},  \emph{Generate PRK fragments}, \emph{Re-encapsulation}, \emph{Decapsulation}, and \emph{Decryption}. These processes are supported by three major cryptographic methods: Key Encapsulation Mechanism (KEM), Data Encapsulation Mechanism (DEM), and Shamir Secret Sharing (SSS)~\cite{10.1145/359168.359176}.

The first step in this process is \emph{Encapsulation}. This is achieved through the use of a Key Encapsulation Mechanism (KEM), in this case, a loosely inspired implementation of ECIES-KEM introduced by~\cite{cryptoeprint:2001/112}. The KEM is fed with Alice's public key $pk_A$ and outputs a symmetric key $K$, and a \emph{capsule}.

With the capsule and the symmetric key $K$, the \emph{Encryption} process is performed using a Data encapsulation mechanism (DEM) which uses Authenticated Encryption with Additional Data (AEAD). This outputs a ciphertext encrypted with the symmetric key.

When the data is encrypted and stored in the cloud, in order for the access delegation to occur, there is a need to generate a PRK. This is performed by the \emph{Generate PRK fragments} process resorting to the notions present in Shamir Secret Sharing (SSS), Alice's private key and signing key $signk_A$, and Bob's public key $pk_B$. This enables the generation of the PRK fragments or \emph{kFrags}. The number of fragments is defined by the number of shares.

The \emph{kFrags} are stored by the proxy for further use in the \emph{Re-encapsulation} process. This process is responsible for generating the \emph{cFrags} which enables Bob to gain access to the file at a later stage. To generate the \emph{cFrags} just the capsule and the \emph{kFrags} are needed. This is due to the fact that this PRE scheme performs the re-encryption over the capsule.

Lastly, once Bob wants to retrieve a file, the \emph{Decapsulation} process needs to happen. This process resorts to SSS in order to reconstruct the symmetric key $k$. To do so, Alice's public key, Alice's verifying key $vk_A$ for signature verification of the \emph{cFrags}, Bob's private key $sk_B$, and the \emph{capsule} are needed. Through the use of a Key Derivation Function within the KEM, it is possible to derive the symmetric key $K$ which together with the ciphertext is passed to the DEM. The DEM performs the \emph{Decryption} process and outputs the plaintext content of the file that Bob can now use.

\section{Related work} \label{sc: related-work}

The notion of PRE made its first appearance in 1998 when Blaze et al.~\cite{blaze1998divertible} introduced the concept of bidirectional and multi-use PRE. Several works have been published since then with new PRE schemes providing new functionalities and relying on different mathematical assumptions. For example, both \cite{10.1007/978-3-642-27954-6_22}  and \cite{pkc-2014-25266} proposed a unidirectional, single-use PRE scheme, but the first relies on threshold PKE, while the second is based on lattice-hardness problems. In 2015, \cite{Liang15} also proposed a unidirectional and single-use PRE scheme, which can be classified as attribute-based. Later, in 2017, ~\cite{Umbral} presented a unidirectional, non-interactive, and verifiable PRE scheme which is threshold-based.

In the context of healthcare data sharing, PRE has also been widely explored. In fact, several works address security, privacy, and confidentiality when it comes to the design and implementation of e-health systems. However, there is still a lack of development concerning safeguarding emergency scenarios in the context of e-health systems~\cite{YUKSEL20171}. Works that address this kind of scenario in its design, refer to this as break-glass approaches.
In 2017, \cite{AU201746} proposed  a framework for the secure sharing of Personal Health Records (PHRs) that relies on attribute-based PRE and which addresses emergency scenarios. The break-glass capabilities are provided with ABE. In this scheme, the emergency department attribute is always appended to the policy that encrypts the patient PHR, thus providing instant access to the entity from the moment the same is uploaded.
The problem with this approach, and in general with ABE approaches, is that they present some caveats, namely key management and resorting to other mechanisms in break-glass approaches. This is due to the fact that emergency normally means an exception to a policy and, thus, overriding that same policy might be a hefty task in some implementations.
In 2019, \cite{YANG2019567} also proposed an approach that is based on an attribute-based PRE, and provided self-adaptive access control, meaning that the system can automatically adapt to normal or emergency situations. However, their break-glass mechanism resorts to a \emph{password-based} paradigm. This approach raises some concerns, namely in the assumption that the individual that stores the password, has the necessary means to ensure its secrecy.
More recently, in 2022,~\cite{LiJinKumari} proposed a system for IoT sensors combining PRE and PKE with equality test, permitting searches under different public keys and secure data sharing. However, it does not discuss emergency situations.  In the same year, \cite{app12094353}, proposed a non-interactive, multi-use, certificateless PRE for sharing health data in a cloud environment. Even though their approach gives full control to the data owner, it has two important drawbacks, namely it is interactive and does not propose a break-glass mechanism. Also in 2022, \cite{9718101} published a secure data sharing and authorised Searchable framework for e-healthcare systems. This framework lies on a conditional and unidirectional PRE scheme with keyword search. It is also idealised for managing sensible data from medical wearable devices. This platform has some disadvantages namely regarding the PRK generation performance. Also, this work does not address emergency situations.
Finally, in 2022, \cite{10.1007/978-3-642-16161-2_6} propose a framework which is also based on attribute-based PRE that features break-glass capabilities. However, it leaves open the possible solution for revocability. 
That being said, there is a need to develop a solution that can cope with all the aforementioned concerns and that contributes to a more reliable and robust break-glass approach.

\section{Patient-centric health data sovereignty} \label{sc: data-sovereignty}
In this section, we introduce the envisioned solution for a patient-centric platform that enables health data sovereignty through PRE. The subsequent section presents the architecture of the solution, followed by a description of the processes involved in the key operations for access delegation.
\subsection{Proposed Solution}

\begin{figure}[h!]
    \includegraphics[width=\textwidth]{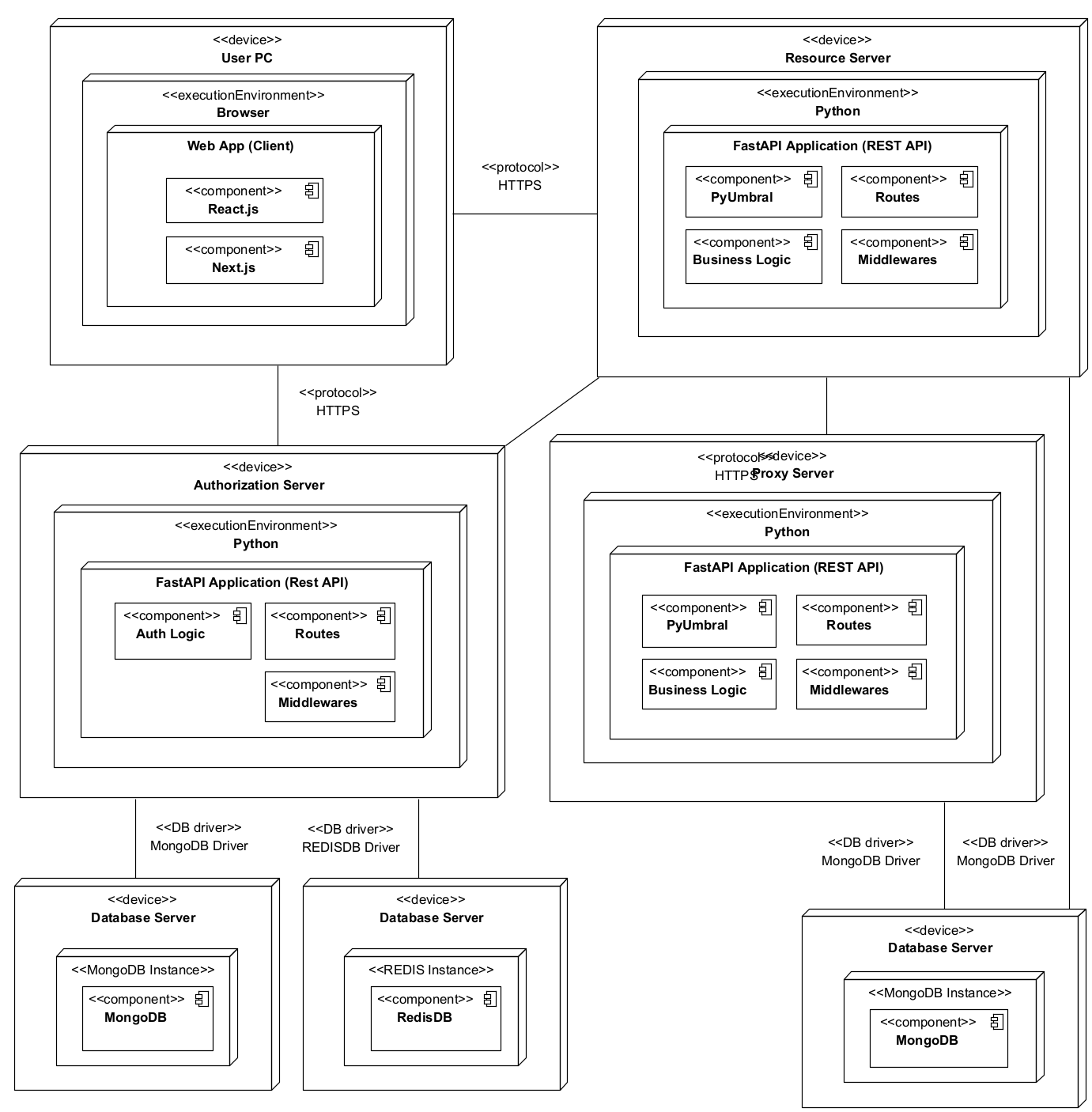}
    \caption{Deployment diagram of the idealised architecture}
    \label{fig: deployment-diagram}
\end{figure}

The proposed solution consists of four main nodes: the client, the resource server, the proxy server, and the authorization server, as depicted in Figure~\ref{fig: deployment-diagram}.

The client node hosts the client-side application developed with Next.js\footnote{https://nextjs.org/}. This client node communicates with the server nodes via Representation State Transfer (REST) and the Hypertext Transfer Protocol (HTTP).
The business logic is divided between the resource and proxy server nodes. The resource server is based on the FastAPI framework\footnote{https://fastapi.tiangolo.com/} running in a Python environment. This server is trusted by the data delegator and it is responsible for assisting the client-side operations, namely feeding the data the client node needs to display the information to the user. The resource server node also performs some core operations such as the initial encryption and final decryption of the Electronic Health Record (EHR) stored in the database server node hosted in a cloud environment (MongoDB\footnote{https://www.mongodb.com/}) as well as the management of delegation requests (accept or decline). Some other complementary operations are also performed such as the generation of the PRK which is stored afterwards by the proxy server node, and the signature verification of the PRK fragments and capsule fragments.

The proxy server is solely responsible for the process of EHR delegation, being used for the re-encryption of the capsules and the storage of the PRK.

The authorisation server is responsible for performing the authentication of the different users of the platform as well as the issuing and claims verification of the authorisation tokens. These tokens are subsequently used to consume the APIs provided by the resource and proxy server nodes. This node is also associated with two persistence nodes. An in-memory database (REDIS\footnote{https://redis.com/} instance) for persisting and performing the lookup of the refresh tokens and a MongoDB instance for storing general purpose user information such as name, email, password, public and verifying keys and roles.

%\subsection{Data modelling} se houver espaço :)

\subsection{Authentication/Authorisation}

The authorisation is performed by resorting to JSON Web Tokens (JWT) which are signed using HMAC SHA256. This ensures the tokens can not be tampered with or changed, thus enabling secure transmissions between parties.

The authentication flow comprises a traditional email/password authentication, where each user needs to provide a valid email and password. In case of successful authentication, a pair of tokens are issued (access/refresh token) containing some claims needed to support the client-side application. These claims follow the standards and restrictions defined in Request for Comments 7519\footnote{https://datatracker.ietf.org/doc/html/rfc7519\#section-4}. Besides the pair of tokens, a Cross-site Request Forgery token is also sent for further protection in requests that require cookies. The refresh token is also sent in a cookie configured with the \emph{secure} and \emph{httpOnly} flags to ensure it is only transmitted through HTTPS and not available to JavaScript in case of a Cross-site Scripting vulnerability in the client-side application.

Since JWT tokens are self-contained, there is no natural way of revoking them. In order to tackle this problem anti-theft mitigation techniques were implemented: \emph{refresh token rotation} and \emph{token reuse detection}.

% \textbf{Refresh token rotation} is achieved by issuing a new pair of tokens every time a new access token is issued. When the access token expires, the client-side application requests a new access token by sending the refresh token. If the refresh token is valid a new pair of tokens is issued and the current refresh token is revoked.

% This revoking of refresh tokens is achieved through the use of a whitelist approach that monitors which tokens are validly issued. Once a refresh token is issued, the same is removed from the whitelist of tokens and thus deemed as expired. This is further motivated by the support of multiple devices. In case of a scenario where a user is authenticated in multiple devices and one of the devices is stolen, the same can use the other device to invalidate all the refresh tokens issued for its account, and thus, limit the access to the stolen refresh token.

% The effectiveness of this solution is further expanded with \textbf{token reuse detection}. By detecting events of reuse of a refresh token, there is an increase in the safety of the platform in general, since it is possible to detect malicious intents regarding a specific user. This is achieved by checking if the sent refresh token has a valid timestamp and signature but is not part of the whitelist of unused refresh tokens. If so, a case of reuse is detected, and thus, the whole family of tokens from a given user is deleted, invalidating all the issued refresh tokens, and forcing the user to re-authenticate.

\subsection{Access delegation scenario} \label{ssc: access-delegation-scenario}

Access delegation is the core problem tackled in this work. The next sections dissect the access delegation flow from the moment the file is uploaded by the patient to the moment the plaintext content is retrieved by the healthcare provider. For demonstration purposes, the step-by-step process between two entities, Alice (delegator) and Bob (delegatee), is presented.

\noindent\textbf{Upload of an EHR} The access delegation starts with the upload of an EHR by Alice. When Alice uploads a new EHR, which can be a Portable
Document Format (PDF) or an image, the resource server encrypts the file using a symmetric key resulting from the encapsulation process and stores it together with the capsule, resulting from the \emph{encapsulation} process, and an associated \emph{userId}.

Another process that is also performed in this step and further detailed in Section~\ref{ssc: break-glass-approach} is the safeguarding of emergency situations. Besides the persistence of the file in the database, a PRK is also generated in order to provide access to the predefined trusted entity. This ensures that the trusted party possesses the means to access the file from the moment it is uploaded and that no extra input from the user is needed in this regard. This PRK is sent to the proxy for subsequent use.

% \begin{figure}[h!]
%     \includegraphics[width=\textwidth]{figures/pre-concrete-sequence-diagram2.png}
%     \caption{Access delegation flow - File Upload}
%     \label{fig: pre-concrete-sequence-diagram-step1}
% \end{figure}

\noindent \textbf{Bob requests access to an EHR} When Bob wants to access Alice's uploaded EHR, he needs to formalise his intentions by issuing a share request to the resource server containing the EHR's \emph{resourceId}. In this step, the system checks if Bob is the owner of the EHR. This prevents a user from performing a share request to itself, something that violates the business rules of the platform.

Once this validation is performed, and provided with the \emph{resourceId}, the resource server generates a share request that includes the \emph{resourceId}, the \emph{delegatorId} and the \emph{delegateeId}, as well as a \emph{status} that is set to pending by default.

% \begin{figure}[h!]
%     \includegraphics[width=\textwidth]{figures/pre-concrete-sequence-diagram3.png}
%     \caption{Access delegation flow - Bob requests access to a file}
%     \label{fig: pre-concrete-sequence-diagram-step2}
% \end{figure}

\noindent \textbf{Alice answers the share request} Now that Bob asked Alice for access to the EHR, Alice is now capable of answering the share request. Depending on Alice's answer, the execution flow might have two outcomes:

\textbf{Accept scenario} --- In case of an acceptance, Alice needs to generate the PRK required to re-encrypt the capsule and further enable Bob to have access to the plaintext content of the EHR. To achieve such a feat, Alice requires his secret key along with his signing key pair, needed to verify the signature of the \emph{kFrags} and \emph{cFrags} at a later stage, as well as Bob's public key. Notice that just the public key is needed, due to the non-interactivity property of this PRE scheme. Lastly, since the underlying scheme of the access delegation mechanism is a threshold PRE scheme, there is also the need to provide a \emph{threshold} which defines the minimum number of shares needed to decrypt the capsule and the number of \emph{shares} which dictates the number of outputted PRK fragments. This last aforementioned operation outputs the \emph{kFrags}, which are sent to the proxy along with a \emph{shareId} binding the PRK to a specific share request. Both attributes are persisted by the proxy for further use once Bob retrieves the EHR.

The share request operation ends with the status update of the share request, which is defined as accepted, together with an arbitrary expiration date defined by Alice. This expiration date is optional, being possible to share an EHR indefinitely or temporarily, in which case the share request is automatically revoked through a cron job once that date is transposed. This ensures the time-based access delegation aspect that this work contributes to.

\textbf{Decline scenario} --- In case Alice declines the share request the status is updated accordingly and no other action is performed.

% \begin{figure}[h!]
%     \includegraphics[width=\textwidth]{figures/pre-concrete-sequence-diagram4.png}
%     \caption{Access delegation flow - Alice answers the share request}
%     \label{fig: pre-concrete-sequence-diagram-step3}
% \end{figure}

\noindent\textbf{Bob retrieves the EHR}
Now that Alice explicitly delegated access to the EHR, Bob is now capable of retrieving it. To do so, Bob performs a request to the resource server, which requires Bob's secret key and the \emph{resourceId}, which uniquely identifies the EHR. A file ownership verification is also performed since the decryption steps are different for a delegator and a delegatee, where the former does  not have the need to re-encrypt the \emph{capsule}.

As stated previously, ownership trails different paths regarding execution flow. With that said, the following can happen whether the user is or not a data owner.

\textbf{Data owner} --- In case the user that requests the file is a data owner, a hybrid encryption approach is used, thus no re-encryption takes place.

\textbf{Not a data owner} --- If the user is not the data owner, meaning they are a delegatee, a collaborative operation between the resource and proxy servers is required to take place. For this specific scenario, Bob needs to ask the proxy to re-encrypt the capsule using the previously generated PRK. To that purpose, the resource server retrieves the EHR details and sends the capsule to the proxy server. The proxy, equipped with the capsule and the PRK fragments \emph{kFrags}, performs the \emph{re-encapsulation} process outputting the \emph{cFrags}. These \emph{cFrags} are sent back to the resource server, which validates their signature through Alice's verifying key.
Once the capsule fragments are validated, Bob decrypts the file by opening the capsule. This last step encompasses Bob's private key, Alice's verifying key and the verified \emph{cFrags}. With the plaintext content of the EHR, Bob is now capable of accessing the information.

% \begin{figure}[h!]
%     \includegraphics[width=\textwidth]{figures/pre-concrete-sequence-diagram5.png}
%     \caption{Access delegation flow - Bob retrieves the file}
%     \label{fig: pre-concrete-sequence-diagram-step4}
% \end{figure}

\textbf{Some important remarks} to highlight are that the secret key used in the sharing process is never shared with the intermediary entity or proxy, making it semi-trusted. Additionally, the proxy only stores the PRK, which alone does not grant it the capability to decrypt the file. Furthermore, even if the stored information such as the capsule, PRK, and ciphertext were to be leaked from the database, the safety and integrity of the EHRs would still be preserved, as they are not sufficient for decrypting the EHRs.

\subsection{Break-glass approach} \label{ssc: break-glass-approach}

Safeguarding emergency scenarios is of paramount importance in a health-related platform. Therefore, we adopted amd approach that features a central trustworthy entity responsible for managing the authorisation in emergency scenarios. This trustworthy entity is seen as a government entity that is responsible for managing such issues and has full access to the files submitted in the platform.

The implementation is similar to what is described in Section~\ref{ssc: access-delegation-scenario} regarding Alice accepting the share request. However, in this case, there is no explicit acceptance of the share request. When an EHR is uploaded, the trusted entity user is retrieved from the database and a PRK is generated. An accepted share request is automatically created for the trusted entity, which links the PRK to the share request between the patient and the trusted entity.

Regarding the process of retrieving the EHR, it follows a similar procedure as depicted in Section~\ref{ssc: access-delegation-scenario}. Just like in a regular file retrieval, since the share request is automatically accepted and the proxy possesses the PRK, the trusted entity requests the proxy to re-encrypt the capsules, enabling the final decryption to take place.

This approach vastly reduces the dependency on external actors, increasing the reliability and availability of the idealised break-glass approach. Having a dedicated entity for this purpose enables instant and swift access to the information if needed.

\section{Performance analysis} \label{sc: performance-analysis}

In this section, we present the performance tests conducted to evaluate our platform. Given the common concerns of limited hardware infrastructures and sub-optimal conditions in governmental adoption cases, it is important to assess the responsiveness of the key operations offered by the platform. Our main goal is to quantitatively analyze the performance of the most computationally intensive operations and assess the impact of the PRE scheme. As there are no specific regulations, indications, or suggestions regarding performance for this type of platform, our tests are purely quantitative and based on known factors and conditions.

The performance tests were carried out on a deployed version of the platform, hosted in Microsoft Azure using a Free F1 tier running Linux and Python 3.10. While these specifications may be basic, they are sufficient to simulate a sub-optimal environment. In real-world scenarios, it is common for governments to have financial restrictions, making it likely that the platform would be deployed on infrastructure with modest specifications. The tests were conducted using Apache JMeter as the tool of choice.

In the rest of this section, we present the results related to the three most crucial operations of the platform and which involve the use of PRE: file upload, accepting a share request, and file retrieval. Additionally, a brief analysis of the results is also presented.

\textbf{File upload}
The performance tests depicted in this section aim to evaluate how the different file sizes impact the upload performance of files.

Since the size of EHRs depends on various factors, such as the patient's medical history, the image resolution of the machines used for exams, and the content of the file itself, determining an average file size becomes challenging. Therefore, we conducted our experiments using two different file sizes: 1MB and 10MB.

\begin{figure}[h!]
    \centering
    \includegraphics[width=0.9\textwidth]{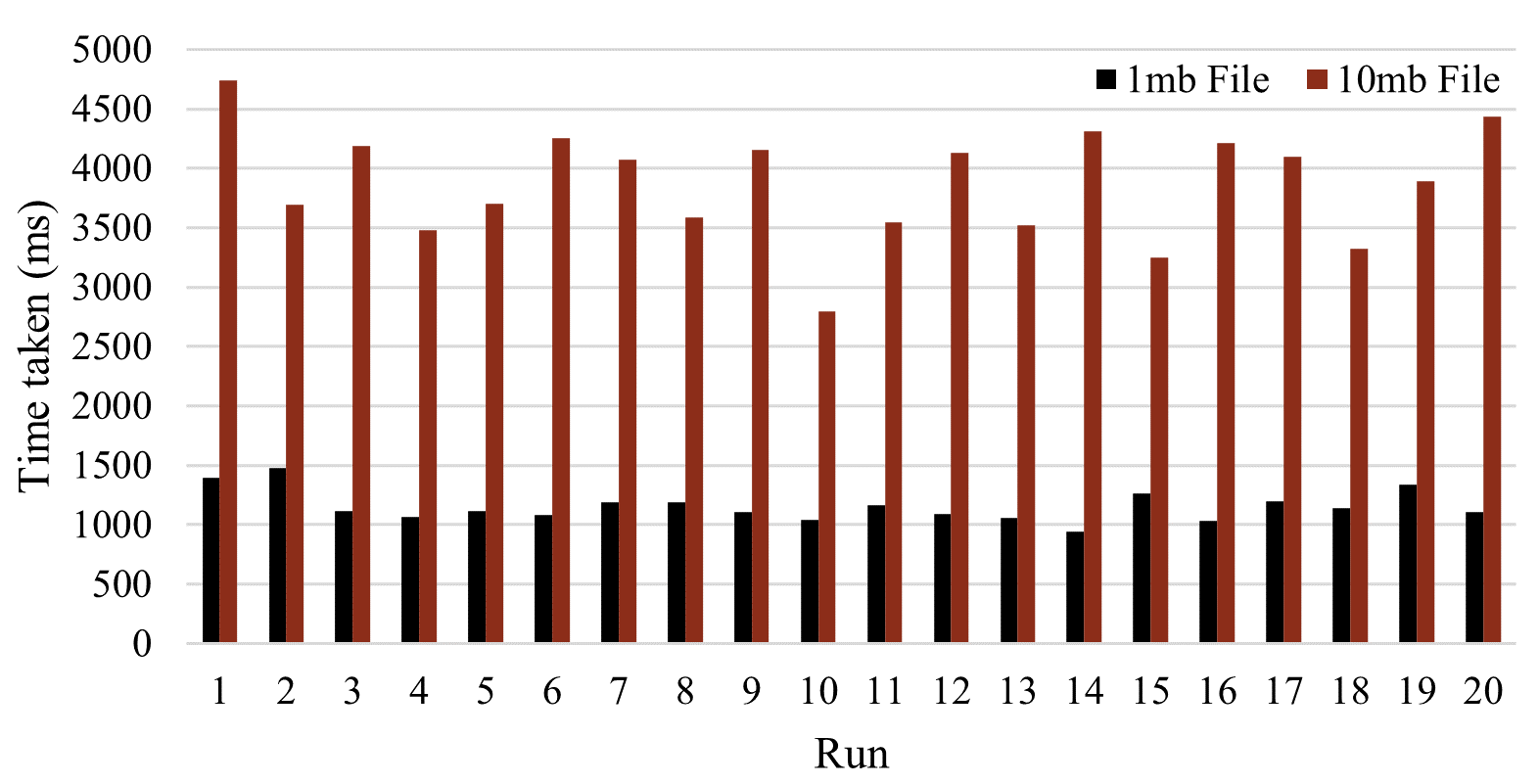}
    \caption{Performance Tests - File Size Uploads Bar Chart}
    \label{fig: performance-tests-file-size-bar-chart}
\end{figure}

Figure~\ref{fig: performance-tests-file-size-bar-chart} illustrates the results obtained from a series of twenty runs performed for each file size.

It can be observed that a tenfold increase in file size reflected an average increase of 2715~\emph{milliseconds}(ms) when comparing file sizes of 1MB and 10MB respectively. The former took an average of 1154~\emph{ms} and the latter an average of 3870~\emph{ms}.

Despite a time of almost four \emph{seconds}, and considering this is not an ideal response time for a REST API, it should be taken into account the complexity of the operations performed. Since this is not a critical operation when it comes to performance, these values are acceptable.

\textbf{Accepting a share request}
The acceptance of a share request is a key operation in the platform described in this paper. Although its performance does not possess a high impact on the efficiency of the platform, it does provide valuable information regarding the PRE process. In this operation, the PRK is generated and sent to the proxy for persistence purposes. Notice that, in this case, there was no need to perform the tests for both file sizes since the PRK generation only depends on cryptographic keys.

Regarding the results of these tests, the average time obtained in 20 runs was 869~\emph{ms}. This quick response was expected since the generation of the PRK fragments is a relatively simple operation that depends on the cryptographic key from both ends, the signature, and the number of shares. Additionally, there was not a significant variation among the twenty runs that were performed. This is supported by the low standard deviation of just 188~\emph{ms}.

% The results presented in Table~\ref{tab: perforamce-tests-time-taken-kFrags} display the average time taken for a PRK with 20 shares and a threshold of 10 shares.

% \begin{table}[h!]
% \centering
% \caption{Performance Tests - Average time taken to generate the PRK}
% \label{tab: perforamce-tests-time-taken-kFrags}
% \begin{tabular}{|c|c|c|}
% \hline
% Runs & Average Time (ms) & Standard Deviation (ms) \\ \hline
% {[}1-20{]}    & 869                        & 188                              \\ \hline
% \end{tabular}
% \end{table}

\textbf{File retrieval} This set of tests aims to assess the impact of file sizes and the use of PRE on a file retrieval scenario. The tests were conducted for both regular decryption and PRE decryption. To evaluate the impact of file sizes, the tests were performed for both 1MB and 10MB file sizes.

\begin{figure}[!h]
    \begin{minipage}[c]{0.49\linewidth}
        \centering
    \includegraphics[width=\textwidth]{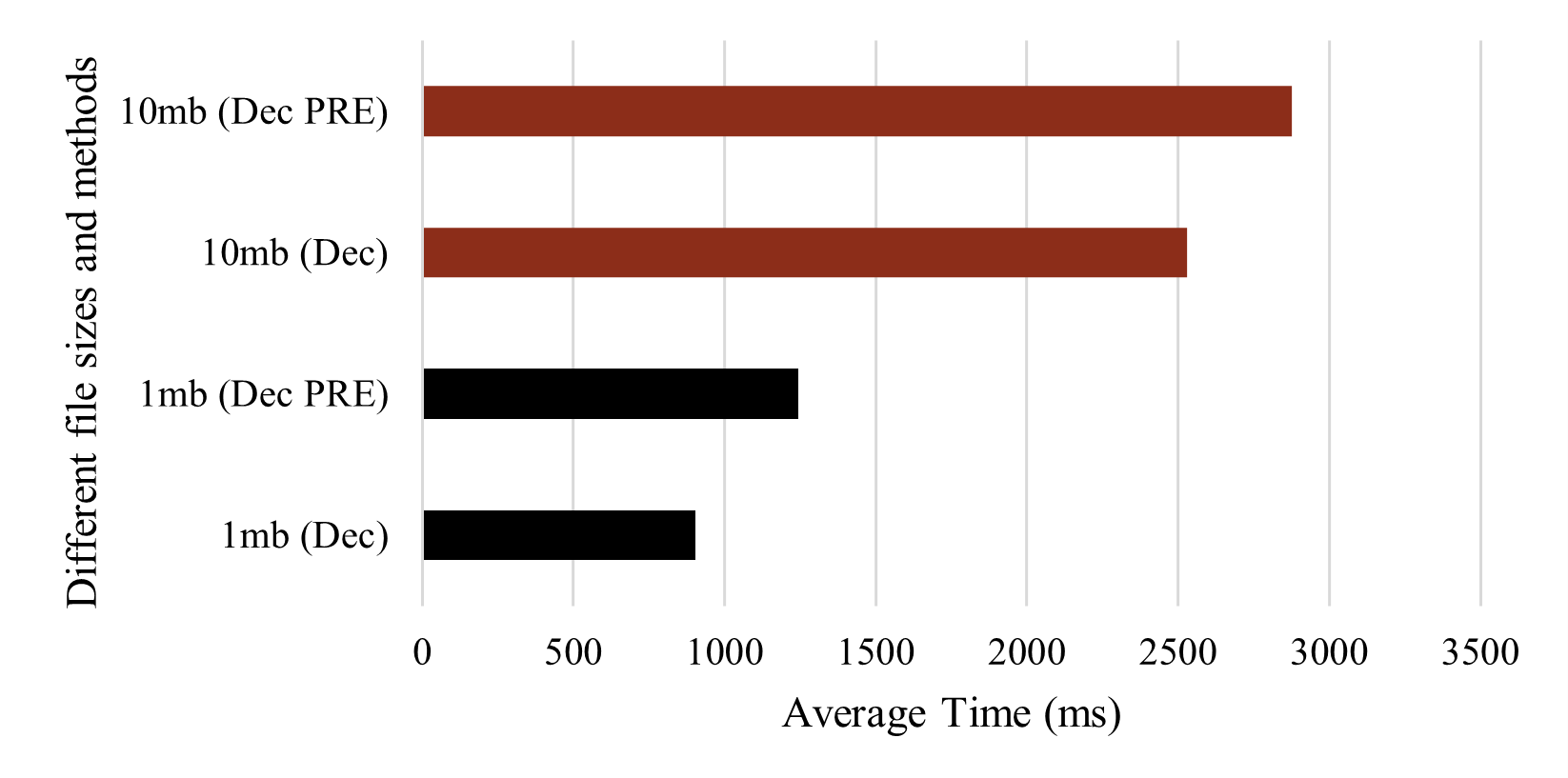}
    \caption{Performance Tests - Average Time Taken for File Retrieval}
    \label{fig: performance-tests-average-time-retrieval-bar-chart}
    \end{minipage}\hfill
    \begin{minipage}[c]{0.49\linewidth}
        \centering
        \includegraphics[width=\textwidth]{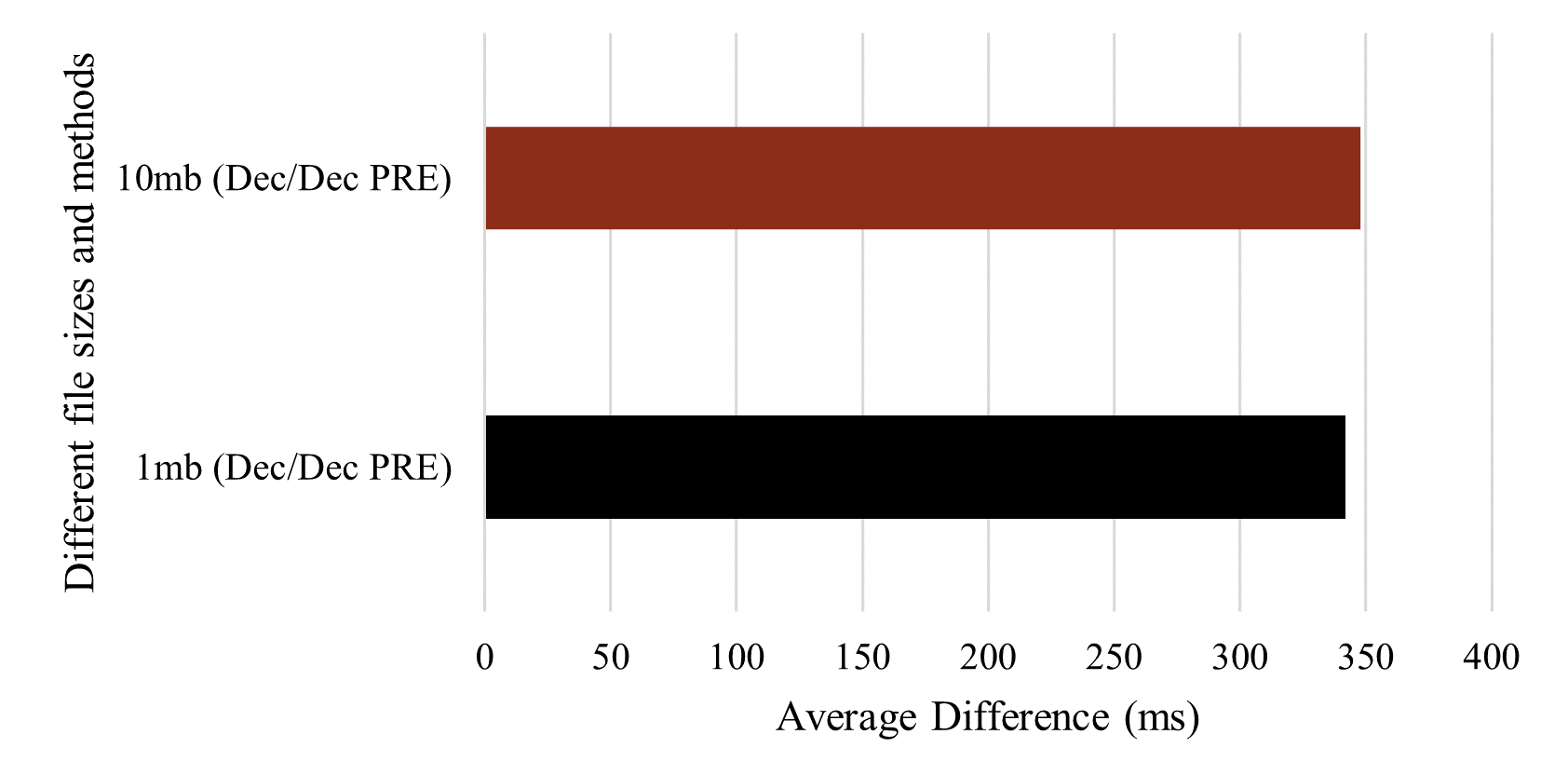}
         \caption{Performance Tests - Average Impact of PRE in Same Sized Files}
         \label{fig: performance-impact-time-retrieval-samesize-bar-chart}
    \end{minipage}
\end{figure}

Moving on to the obtained results (Fig.\ref{fig: performance-tests-average-time-retrieval-bar-chart}), a 1MB file took an average of 903~\emph{ms} to be retrieved while the 10MB one took an average of 2529~\emph{ms}. Regarding file retrieval with PRE, the 1MB file took an average of 1245~\emph{ms} and 2877~\emph{ms} for the 10MB file.

We have also evaluated the impact of re-encryption on file retrieval operations (Fig.\ref{fig: performance-impact-time-retrieval-samesize-bar-chart}) by directly measuring the difference between regular decryption and PRE for each file size. This resulted in an average difference of 342~\emph{ms} for the 1MB file and 348~\emph{ms} for the other one.

The results of our tests indicate that there is a similar average difference between regular and PRE decryption for both file sizes. This similarity can be attributed to the fact that the re-encryption process only affects the capsule, not the actual file. Since the sizes of the capsule and cryptographic keys are similar in both scenarios, it is expected that the results would be similar as well. The file size does not significantly impact the re-encryption of the capsule, but rather affects the overhead associated with fetching the file from the database and delivering it in the response.

% \begin{figure}[h!]
%     \centering
%     \includegraphics[width=0.8\textwidth]{figures/average-impact-different-filesizes.png}
%     \caption{Performance Tests - Average Impact of PRE in Different Sized Files}
%     \label{fig: performance-impact-time-retrieval-different-size-bar-chart}
% \end{figure}

Regarding the obtained results, they were deemed satisfactory since most operations do not possess restrictive requirements when it comes to performance.

Regarding more critical operations such as file retrieval, considering the computational effort and infrastructure complexity required to ensure full correctness with the underlying threshold PRE scheme, the results were deemed satisfactory. It is important to note that these tests were conducted in a shared infrastructure with modest specifications. Thus, it was not possible to control the current workload of the servers during the tests, which may have impacted negatively the aforementioned results

\section{Conclusion} \label{sc: conclusion}

In this paper, we present a PRE-based platform for the secure sharing of e-health, considering a sovereign approach focused on the patient. This approach is achieved by ensuring that the patient's data is only shared with their explicit consent. Furthermore, it also enables robust revocability by the patient, without requiring updates on the encrypted EHR database, further contributing to a user-centric approach. Non-interactivity is also a key characteristic of our platform, which does not require sharing user's private key for the re-encryption process to occur. Another key achievement of our work is the proposed break-glass mechanism. Since some implementations fall short in terms of revocability, and only a few contemplate PRE in emergency scenarios, our solution uses a central trusted entity to which the proxy delegates access from the moment the EHR is uploaded to the platform. This eliminates the need to trust external actors in the system, increasing reliability and allowing swift access to the information in critical situations.
There are other key characteristics of our platform worth highlighting. Firstly, it uses symmetric encryption to encrypt the EHR, which is faster than PKE. Secondly, the re-encryption process is performed over the capsule, which tends to have a much smaller size compared to a PHR. 
The tests that were conducted and our results show that the most demanding task is the upload of the EHR, as expected, because it requires the encapsulation process to occur and the encryption of the EHR. However, the re-encryption process does not show a significant increase when the size of the uploaded files increases. This is because the re-encryption does not involve the EHR.
Our platform provides a solution to the sharing of medical data that incorporates key functionalities not covered together in previous literature, such as unidirectionality, non-interactivity, revocability, and a mechanism to deal with emergency situations. This solution contributes to the collaborative aspect of e-health and enables better, and more informed treatments supported by the increased exchange of information between providers.

Regarding future work, it would be beneficial to extend the architecture to accommodate multiple proxies instead of using just one. This could be achieved by utilising a blockchain network where the proxies work together to re-encrypt the capsules, thus enabling all the benefits that a threshold-based scheme has to offer. Furthermore, additional tests could be performed using different environments and network conditions to cover more use case scenarios.

\bibliographystyle{splncs04}
\bibliography{BIB}

\end{document}